\documentclass{article}

\PassOptionsToPackage{numbers, compress}{natbib}

\usepackage{multirow}

\usepackage[final]{neurips_2024_ml4ps}

\usepackage[utf8]{inputenc} 
\usepackage[T1]{fontenc}    
\usepackage{hyperref}       
\usepackage{url}            
\usepackage{booktabs}       
\usepackage{amsfonts}       
\usepackage{nicefrac}       
\usepackage{microtype}      
\usepackage{xcolor}         
\usepackage[nolist,nohyperlinks]{acronym}

\usepackage{mathtools}
\usepackage{soul}
\newcommand*\diff{\mathop{}\!\mathrm{d}}

\definecolor{peach}{HTML}{F26035}

\usepackage{amsmath}
\usepackage[caption=false]{subfig}

\newcommand{\prob}{\ensuremath{{p}}}

\newcommand{\data}{\ensuremath{{d}}}

\newcommand{\given}{\ensuremath{{\,\vert\,}}}

\title{Accelerated Bayesian parameter estimation and model selection for gravitational waves with normalizing flows}

\author{%
Alicja Polanska \\
Mullard Space Science Laboratory\\
University College London\\
  \texttt{alicja.polanska.22@ucl.ac.uk} \\
  \And
  Thibeau Wouters\\
  Utrecht University \\
  Nikhef \\
  \texttt{t.r.i.wouters@uu.nl} \\
  \AND
  Peter T. H. Pang\\
  Nikhef \\
  Utrecht University \\
  \texttt{thopang@nikhef.nl} \\
  \And
  Kaze K. W. Wong \\
  Center for Computational Astrophysics \\ 
  Flatiron Institute \\
  Department of Applied Mathematics and Statistics \\
  Johns Hopkins University \\
  \texttt{wwong24@jhu.edu} \\
  \And
  Jason D. McEwen \\
  Mullard Space Science Laboratory\\
  University College London\\
  Alan Turing Institute \\
  \texttt{jason.mcewen@ucl.ac.uk} \\
}

\begin{document}

\maketitle

\begin{abstract}
    We present an accelerated pipeline, based on high-performance computing techniques and normalizing flows, for joint Bayesian parameter estimation and model selection and demonstrate its efficiency in gravitational wave astrophysics. We integrate the \textsc{Jim} inference toolkit, a normalizing flow-enhanced \ac{MCMC} sampler, with the learned harmonic mean estimator. Our Bayesian evidence estimates run on $1$ GPU are consistent with traditional nested sampling techniques run on $16$ CPU cores, while reducing the computation time by factors of $5\times$ and $15\times$ for $4$-dimensional and $11$-dimensional gravitational wave inference problems, respectively. Our code is available in well-tested and thoroughly documented open-source packages, ensuring accessibility and reproducibility for the wider research community.
\end{abstract}

\section{Introduction}

In many scientific fields Bayesian inference is an indispensable tool for extracting new knowledge from observations, providing a principled statistical framework for parameter estimation and model selection.  In Bayesian statistics, the posterior probability distribution $\prob(\theta \given \data, M)$ encodes information about the parameters $\theta$, of model $M$, given the observed data $\data$. By Bayes' theorem the posterior is given by
\begin{equation}\label{eq:bayes}
    \prob(\theta \given \data, M)
    = \frac{\prob(\data \given \theta, M) \prob(\theta \given M)}{\prob(\data \given M)},
\end{equation}
where $\prob(\data \given \theta, M)$ is the likelihood, $\prob(\theta \given M)$ the prior and $\prob(\data \given M)\equiv z$ the Bayesian evidence. The likelihood quantifies how well a model and a given set of parameters $\theta$ describe the data. The prior reflects our existing beliefs about the parameters.  Typically, Markov chain Monte Carlo (MCMC) sampling techniques are used to explore the posterior distribution for parameter estimation, from which parameter estimates and their uncertainties can be computed. The Bayesian evidence $z$, also called the marginal likelihood, is a normalization factor and is computed as
\begin{equation}
    \label{eqn:evidence}
    z = \prob(\data \given M)
    = \int d \theta \, \prob(\data \given \theta, M) \prob(\theta \given M).
\end{equation}
The evidence is a crucial quantity for comparing competing models, allowing us to provide a statistically principled preference for one model over another \citep{Trotta07}; although, it is computationally difficult to calculate.

One particular scientific field that heavily relies on Bayesian inference is \ac{GW} astrophysics. Since 2015, Advanced LIGO~\citep{LIGOScientific:2014pky} and Advanced Virgo~\citep{VIRGO:2014yos} have detected $\mathcal{O}(100)$ \ac{GW} signals originating from  mergers of black holes and neutron stars~\citep{LIGOScientific:2018mvr, LIGOScientific:2020ibl, KAGRA:2021vkt, LIGOScientific:2021usb}, revealing a novel and reliable way of observing and studying the Universe. Evidence estimates have allowed, for instance, one to identify the nature of sources of \acp{GW}~\citep{LIGOScientific:2024elc} and discern between various waveform models encoding different underlying physics~\citep{LIGOScientific:2019eut}, thereby advancing our understanding of \ac{GW} sources. 

Nested sampling algorithms~\citep{Skilling:2006gxv, ashton2022nested} are widely used to compute the Bayesian evidence. However, this method of estimation is tightly coupled to the sampling strategy, inhibiting the adoption of accelerated sampling techniques. The sampling must be performed in a nested manner, which means these methods can be computationally expensive, especially for high-dimensional parameter spaces and multimodal posteriors.  A fast and scalable alternative is therefore of paramount importance for various scientific disciplines. In \ac{GW} astrophysics, for instance, telescope operators require information regarding the nature of the source in low latency to identify potential electromagnetic counterparts of \ac{GW} events.
Moreover, next-generation \ac{GW} detectors, such as the Einstein Telescope~\citep{Punturo:2010zza} and the Cosmic Explorer~\citep{Evans:2021gyd}, will have increased sensitivities which result in longer signal durations and more events to analyze, enhancing the demand for efficient inference methods~\citep{Branchesi:2023mws}. Previous attempts have accelerated nested sampling algorithms using machine learning~\citep{Williams:2021qyt,Williams:2023ppp} or make use of simulation-based inference~\citep{Gabbard:2019rde, Kolmus:2021buf, Kolmus:2024scm, Chua:2019wwt, Green:2020hst, Green:2020dnx, Dax:2021tsq, Dax:2022pxd, Bhardwaj:2023xph}. 

Recently, the \textsc{Jim}\footnote{\url{https://github.com/kazewong/jim}} inference toolkit~\citep{Wong:2023lgb} was introduced, which accelerates parameter estimation by using normalizing flow-enhanced \ac{MCMC} sampling as well as hardware accelerators such as \acp{GPU} and \acp{TPU}. However, \ac{MCMC} methods like \textsc{Jim} do not provide the Bayesian evidence, which is necessary for Bayesian model selection. In this work, we augment \textsc{Jim} with a scalable evidence estimator decoupled from the sampling method -- the learned harmonic mean estimator with normalizing flows~\citep{mcewen2021machine,polanska2024learned}, implemented in the \texttt{harmonic} Python package\footnote{\url{https://github.com/astro-informatics/harmonic}}. Since, unlike nested sampling, the learned harmonic mean is agnostic to the sampling strategy, it is possible to realise the acceleration provided by \textsc{Jim} and still perform accurate evidence estimation. Other methods of evidence estimation decoupled from the sampling strategy have been recently proposed \citep{heavens2017marginal,jia2020normalizing,srinivasan2024floz}, but we choose the learned harmonic mean due to several advantages discussed in depth in Refs. \citep{polanska2024learned,piras2024future}. We demonstrate, using an example from the field of \ac{GW} astrophysics, that our pipeline provides accurate evidence estimates while only requiring a fraction of the computational cost required by the traditional methods.

\section{Methodology}

We construct an accelerated pipeline to, first, sample the posterior distribution and, second, compute the Bayesian evidence. We leverage normalizing flows, at both the sampling and evidence estimation stages. Moreover, we use a sampler that leverages the high-performance computing techniques of \textsc{JAX}~\citep{frostig2018compiling}.

\subsection{Normalizing flows}
Normalizing flows are generative models that transform a simple base distribution into a complex one through a series of invertible, differentiable mappings with learned parameters. The flow can be trained on samples from the distribution of interest by minimising the forward Kullback-Leibler (KL) divergence. For a more extensive review of normalizing flows we refer the reader to references~\citep{papamakarios2021normalizing, Kobyzev_2021}. Both \textsc{Jim} and \texttt{harmonic} use rational-quadratic spline flows~\citep{durkan2019neural}, where piecewise rational-quadratic functions are used in the transformations. They are able to encode nonlinear and local relationships, allowing for a more expressive and powerful architecture than affine transformations~\citep{dinh2016density}.

\subsection{\textsc{Jim} inference toolkit}
In Ref.~\citep{Wong:2023lgb}, the authors introduced \textsc{Jim}, an inference toolkit implemented in \textsc{jax}~\citep{frostig2018compiling}, and applied it to \ac{GW} astrophysics as an example. \textsc{Jim} supports GPU-accelerated differentiable gravitational waveform models~\citep{Edwards:2023sak} and can therefore make use of efficient gradient-based samplers such as the Metropolis-adjusted Langevin algorithm~\citep{grenander1994representations} or Hamiltonian Monte Carlo~\citep{Betancourt:2017ebh}.
In order to further accelerate the parameter estimation, \textsc{Jim} makes use of \textsc{flowMC}\footnote{\url{https://github.com/kazewong/flowMC}} , a normalizing flow-enhanced \ac{MCMC} sampler implemented in \textsc{jax}~\citep{Wong:2022xvh, Gabrie:2021tlu}. It accelerates traditional \ac{MCMC} by adapting a global proposal density distribution to the target distribution with normalizing flow on the fly.
It has been shown that \textsc{Jim} can accurately infer the parameters of \ac{GW} signals originating from merging black holes~\citep{Wong:2023lgb} and neutron stars~\citep{Wouters:2024oxj}. Moreover, \textsc{Jim} achieves this at a fraction of the computational cost of conventional methods that nested sampling, e.g.\ \textsc{Bilby}~\citep{Ashton:2018jfp,Romero-Shaw:2020owr,Smith:2019ucc}.  However, previously \textsc{Jim} could not provide evidence estimates for model comparison.

\subsection{Learned harmonic mean estimator}

To compute the Bayesian evidence from posterior samples, we consider the recently proposed learned harmonic mean estimator \citep{mcewen2021machine,spurio2023bayesian, polanska2024learned, piras2024future}. The learned harmonic mean is a scalable estimator of the evidence based on posterior samples, which is therefore agnostic to the sampler used and can be integrated with the \textsc{flowMC} sampler used in \textsc{Jim}. While the original harmonic mean estimator \citep{newton1994approximate} suffered from instability \citep{neal:1994}, the learned harmonic mean solves this issue by leveraging machine learning techniques \citep{mcewen2021machine}. The reciprocal evidence $\rho = z^{-1}$ is estimated as
\begin{equation}
    \label{eqn:harmonic_mean_retargeted}
    \hat{\rho} =
    \frac{1}{N} \sum_{i=1}^N
    \frac{\varphi(\theta_i)}{\prob(\data \given \theta_i, M) \prob(\theta_i \given M)} ,
    \quad
    \theta_i \sim \prob(\theta | \data, M),
\end{equation}
where $N$ is the number of samples and $\varphi(\theta)$ is a learned normalized target distribution that must be concentrated within the posterior. Recently, the authors of Ref.~\citep{polanska2024learned} integrated normalizing flows into the learned harmonic mean estimator, which provide a robust approach to ensure the learned target distribution is indeed concentrated within the posterior.  Specifically, a temperature parameter is introduced to scale the variance of the base distribution of the flow by a factor $0 < T < 1$. The concentrated flow is then used as the target $\varphi(\theta)$. The authors show that the estimates are robust to different values of $T$. The method is implemented in the \texttt{harmonic} package written in \textsc{JAX}. 

The posterior distributions encountered in \ac{GW} physics are often multimodal, which can prove challenging. In particular, due to the topology-preserving nature of their transformations, normalizing flows tend to struggle with multimodality \citep{cornish2020relaxing}, and are prone to mode-covering behaviour when trained using forward KL divergence \citep{pml1Book}. These problems are mitigated by the fact that for the learned harmonic mean estimator to be accurate, it is not necessary to achieve a very close approximation of the posterior \citep{mcewen2021machine}. However, a poor approximation can potentially be problematic if it leads to regions of high flow density in regions where posterior density is low. To facilitate the learning of the multimodal distributions, we consider a multimodal base distribution (a sum of normal distributions with an identity covariance matrix), which leads to an improvement in \texttt{harmonic} diagnostics. In future work we plan to investigate in detail our method's robustness to multimodality, including for \ac{GW} events with a low signal to noise ratio, by numerically studying the influence of the base distribution choice, as well as considering other flow approaches designed to deal with multimodality \citep[e.g.][]{ziegler2019latent,dinh2019rad, cornish2020relaxing,stimper2022resampling}.

\begin{table}
    \caption{Total wall times to compute the evidence estimates for the examples discussed in the main text. We run \textsc{Bilby} on $16$ CPU cores and \textsc{Jim} + \texttt{harmonic} on $1$ GPU.}
    \label{table:evidences}
    \centering
    \begin{tabular}{c cccc}
        \toprule\toprule
        Example              & Method                           & $\log(z)$                                       & Sampling time                   & Evidence estimation time \\ \midrule
        \multirow{2}{*}{4D}  & \textsc{Bilby}                   & $390.33 \pm 0.11$                               & $31.3$ min                      & --                       \\
                             & \textsc{Jim} + \texttt{harmonic} & $390.360^{+0.006}_{-0.006}$                     & $\phantom{0}3.4$ min            & $1.9$ min                \\ \midrule
        \multirow{2}{*}{11D} & \textsc{Bilby}                   & $378.29 \pm 0.15$                               & $\phantom{0}3.5$ h\phantom{in } & --                       \\
                             & \textsc{Jim} + \texttt{harmonic} & $378.420^{+0.09\phantom{0}}_{-0.08\phantom{0}}$ & $11.8$ min                      & $2.4$ min                \\
        \bottomrule\bottomrule
    \end{tabular}
    \vspace{-1em}
\end{table}

\begin{figure}
    \centering
    \subfloat[\textsc{Bilby}]{\includegraphics[width=0.4\linewidth]{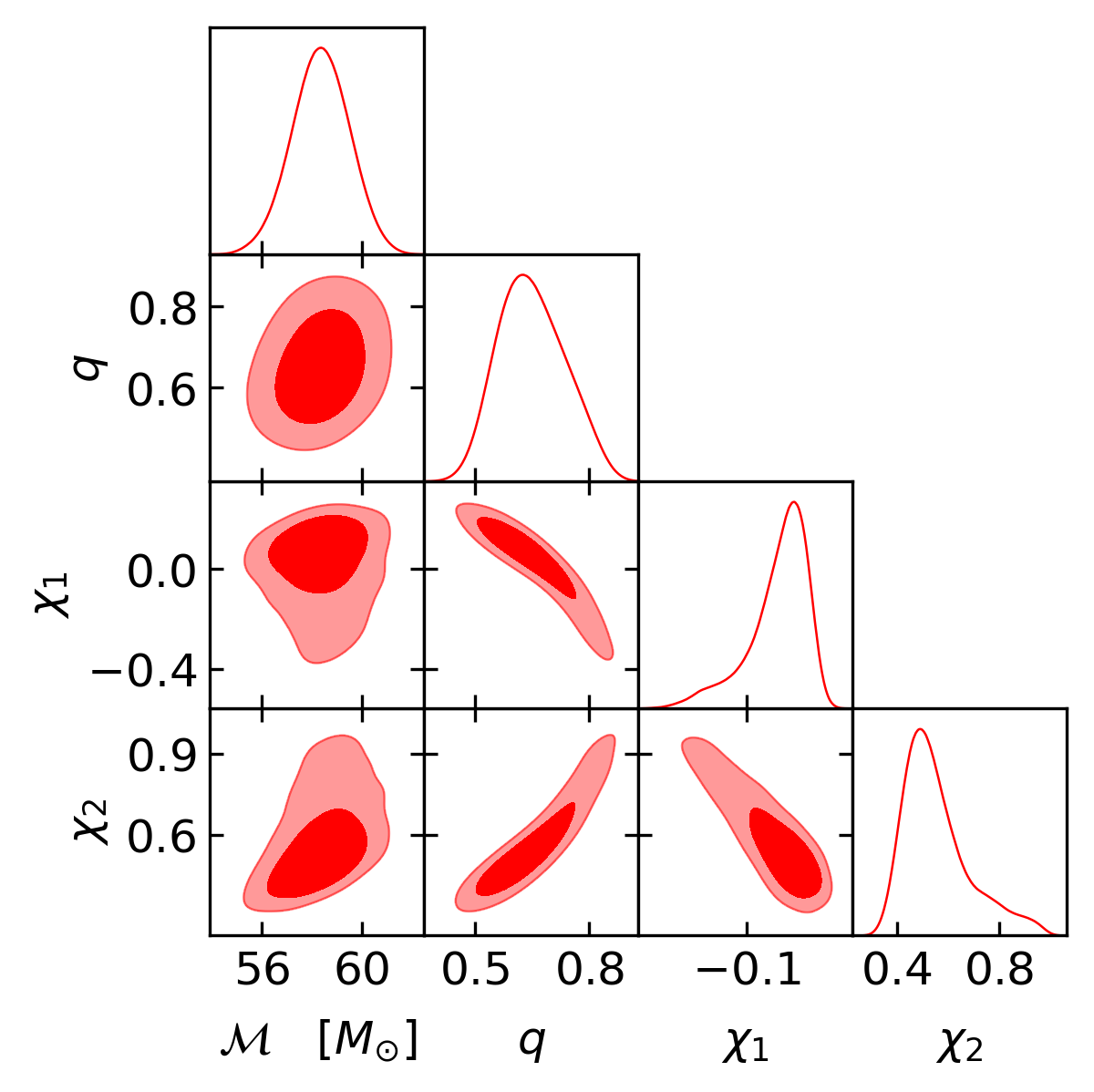}}
    \qquad
    \subfloat[\textsc{Jim} + \texttt{harmonic}]{\includegraphics[width=0.4\linewidth]{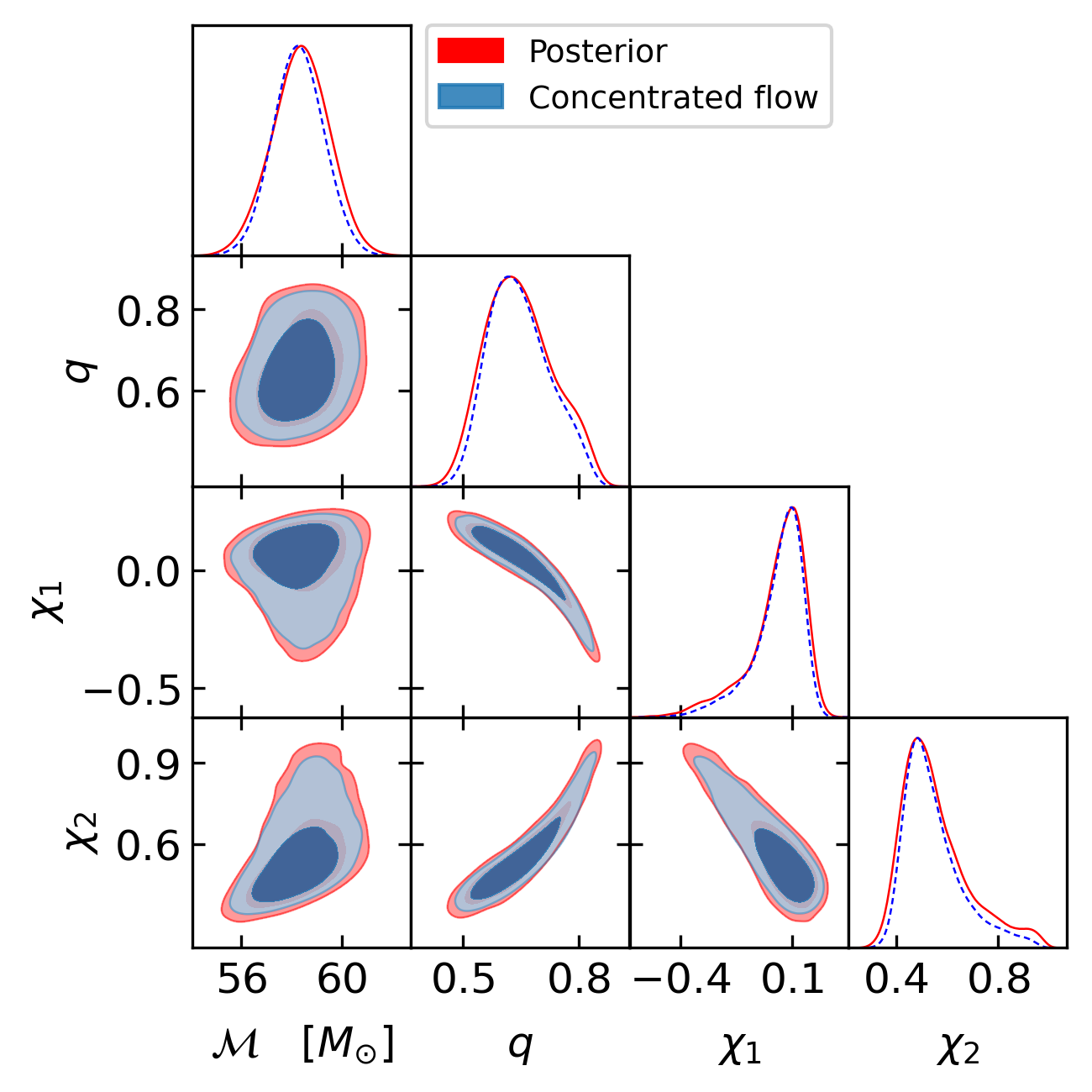}}
    \caption{Corner plots for the 4-dimensional posterior samples from (a) \textsc{Bilby} and (b) \textsc{Jim} used for inference (solid red) alongside the concentrated flow at $T=0.8$ used in the learned harmonic mean (dashed blue). \vspace{-1.8em}}
    \label{fig:4D}
\end{figure}

\section{Results}
\label{sec:results}
We validate and benchmark our pipeline, combining \textsc{Jim} with \texttt{harmonic}, by computing the Bayesian evidence of a simulated \ac{GW} signal from a binary black hole merger as an example.
The signal is injected into a realization of a Gaussian noise time series coming from a network of the two Advanced LIGO~\citep{LIGOScientific:2014pky} and the Advanced Virgo detector~\citep{VIRGO:2014yos} at their design sensitivities.
The recovery of the injected signal is performed in two examples.
In the first example, we only recover the $4$-dimensional intrinsic parameter space, comprising of the masses and the aligned spins of the systems, while fixing other parameters to their injected values.
In the second example, we recover the full $11$-dimensional parameter space, which includes the extrinsic parameters. The setup of these inferences in described in detail in Appendix~\ref{sec:appendix GW setup}. 

For both examples, we compute the Bayesian evidence with nested sampling, and with \textsc{Jim} combined with \texttt{harmonic}.
For the former, we use \textsc{Bilby}~\citep{Ashton:2018jfp}, employing \textsc{dynesty}~\citep{Speagle:2019ivv} as the nested sampling library.
More specifically, we use \textsc{parallel-Bilby}~\citep{Smith:2019ucc} and parallelize the computation with $1000$ live points over $16$ cores on a single Intel Xeon Silver $4310$ Processor \ac{CPU}.
For \textsc{Jim} and \texttt{harmonic}, we use a single NVIDIA A100-$40$GB \ac{GPU} to perform the inference.

When estimating the evidence with \texttt{harmonic} from the \textsc{Jim} posterior samples, we divide the samples into two sets with an equal number of chains, using one to train the flow and the other to estimate the evidence. The samples are thinned, keeping only every tenth sample in each chain, which achieved an accurate estimate at reduced time and memory demands.

As an additional check we also estimate the evidence on posterior samples obtained from nested samples via rejection sampling with \textsc{Bilby}. We randomly shuffle these posterior samples, as the output contains more samples from higher density regions towards the end, introducing bias into the train--inference split. Then we divide them into $20$ chains before using \texttt{harmonic}. We obtain results consistent with nested sampling evidence estimates.

The evidence estimates for the $4$-dimensional example along with their computation times are shown in the first two rows of Table \ref{table:evidences}. We use a rational-quadratic spline flow with $6$ layers and $8$ spline bins and a unimodal base in the learned harmonic mean estimator, and set the temperature parameter to $T=0.8$. We find that the evidence estimates obtained using \textsc{Bilby} are in close agreement with our estimates obtained from \textsc{Jim} samples with \texttt{harmonic}. However, our pipeline achieves a speedup factor of $5.4\times$ relative to \textsc{Bilby} at performing this calculation even for this relatively low dimensional example. 
Figure~\ref{fig:4D} shows corner plots of \textsc{Bilby} samples as as well as the half of \textsc{Jim} samples used for inference, alongside the concentrated flow of \texttt{harmonic}. The plot shows visual agreement between \textsc{Jim} and \textsc{Bilby}, and demonstrates that the flow of \texttt{harmonic} is concentrated in the posterior of \textsc{Jim}. We perform additional sanity checks, described in~\citep{mcewen2021machine} to further validate the results of the learned harmonic mean. In particular we inspect the estimates of error, kurtosis and the ratio between the square root variance of variance and variance estimates.

We repeat the same procedure for the $11$-dimensional example. We use the same thinning procedure and employ $5$ layers with $64$ bins for the rational-quadratic spline flow. Because the multimodal features are more pronounced in this posterior, we use a multimodal base consisting of three normal distributions with an identity covariance matrix, one centered at $0$ in all dimensions, one centered at $0$ except for dimensions $\phi_c$, $\psi$ centered at $1$, and finally one centered at $0$ except for $\phi_c$, $\psi$, $\alpha$, $\delta$ centered at $2$. This heuristic choice introduces the underlying multimodality into the flow at the start of the training and results in improved diagnostics. The results of this analysis are shown in the last two rows of Table~\ref{table:evidences}, with the corner plots shown in Appendix~\ref{sec:appendix_11D_fig}. Evidence estimates are again in close agreement. However, our pipeline is $14.8\times$ faster than \textsc{Bilby}.

\section{Conclusions}

In this work we have constructed an end-to-end pipeline that accelerates Bayesian inference, including both parameter estimation and also model selection.  Our pipeline combines the efficient \ac{MCMC} sampling of \textsc{Jim} with the learned harmonic mean estimator implemented in \texttt{harmonic} to compute the evidence. To demonstrate the effectiveness of our pipeline, we applied it to a simulated \ac{GW} event and inferred its $4$-dimensional intrinsic and complete $11$-dimensional parameter spaces. We have shown that our pipeline provides accurate evidence estimates at only a fraction of the time required by traditional methods, with a speedup of $5.4\times$ and $14.8\times$ for the $4$- and $11$-dimensional examples respectively. In future work, we aim to further investigate the optimal treatment of multimodal distributions, exploring the various approaches proposed in literature \citep[e.g.][]{ziegler2019latent,dinh2019rad, cornish2020relaxing,stimper2022resampling}, and apply the methods presented here to real data of observed \ac{GW} events. Both \textsc{Jim} and \texttt{harmonic} are open-source, well-documented and tested.
Therefore, the pipeline introduced in this work can directly be applied in other scientific fields relying on Bayesian inference. 

\begin{ack}
    AP is supported by the UCL Centre for Doctoral Training in Data Intensive Science (STFC grant number ST/W00674X/1).  TW and PTHP are supported by the research program of the Netherlands Organization for Scientific Research (NWO). JDM is supported by EPSRC (grant number EP/W007673/1) and STFC (grant number ST/W001136/1).  This work was supported by collaborative visits funded by the Cosmology and Astroparticle Student and Postdoc Exchange Network (CASPEN), as well as a G-Research grant. The authors acknowledge the computational resources provided by the LIGO Laboratory's CIT cluster, which is supported by National Science Foundation Grants PHY-0757058 and PHY0823459.
\end{ack}

\medskip

{
    \small

    \bibliography{refs}
}

\appendix

\newpage
\section{Setup of simulated gravitational wave signal}\label{sec:appendix GW setup}

The source parameters $\boldsymbol{\theta}$ of a \ac{GW} signal are inferred from the data by computing their posterior distributions. We set the duration of the simulated signal considered in this work to $4$ seconds and analyze the signal in the frequency domain, setting the frequency range to $[20, 2048]$ Hz. We use the \ac{GW} approximant \texttt{IMRPhenomD}~\citep{Husa:2015iqa, Khan:2015jqa} for injecting and analysing the signal.

In Tab.~\ref{tab:parameter_priors} below, we provide the parameters used in the \ac{GW} simulations and the values chosen for the simulated signal.
All priors used in the analyses are uniform priors in the ranges shown in Tab.~\ref{tab:parameter_priors}.
We adopt the standard convention that $m_1$ refers to the heavier black hole and $m_2$ to the lighter black hole of the binary system such that the mass ratio $q = m_2/m_1$ is bounded above by $1$.
When given a \ac{GW}, represented as time series $d(t)$ and a gravitational waveform model $h(t; \boldsymbol{\theta})$, with $\theta$ the parameters of Tab.~\ref{tab:parameter_priors}, the likelihood function is given by
\begin{equation}
    \begin{aligned}
        \log \prob(\data \given \theta, M) & = - \frac12 \left\langle d - h(\boldsymbol{\theta}), d - h(\boldsymbol{\theta}) \right\rangle + {\rm normalization \ constant}                                                               \\
                                           & = \langle d, h(\boldsymbol{\theta}) \rangle - \frac{1}{2}\langle h(\boldsymbol{\theta}), h(\boldsymbol{\theta}) \rangle - \frac{1}{2}\langle d , d \rangle + {\rm normalization \ constant}.
    \end{aligned}
\end{equation}
Here, the noise-weighted inner product $\langle a, b \rangle$ is defined as
\begin{equation}
    \label{eq:likelihood}
    \langle a, b \rangle = 4 \textrm{Re} \int_{f_{\rm low}}^{f_{\rm high}} \diff f \frac{\tilde{a}(f) \tilde{b}^*(f)}{S_n(f)} \, ,
\end{equation}
with $S_n(f)$ representing the one-sided \ac{PSD}, the asterisk symbol denoting complex conjugation and $\tilde{x}(f)$ representing the Fourier transform of a time series $x(t)$.

As shown in Eq.~\ref{eq:likelihood}, the $\langle d, d \rangle$ term and the normalization constant do not depend on the source parameters $\boldsymbol{\theta}$, which is often neglected in analyses, e.g., in \textsc{Jim}. The Bayesian evidence obtained with such convention is then equivalent to the Bayes factor of the signal hypothesis against the noise hypothesis, thus the Bayes factor quoted by \textsc{Bilby}.

\begin{table*}
    \centering
    \renewcommand{\arraystretch}{1.25}
    \begin{tabular*}{1.0\linewidth}{@{\extracolsep{\fill}} l l l l}
        \hline \hline
        Parameter & Description & Injected value & Prior \\ \hline
        $\mathcal{M}$ & detector-frame chirp mass $[M_{\odot}]$ & $60$ & $[25, 100]$ \\
        $q$ & mass ratio $m_2/m_1$ & $0.65$ & $[0.125, 1]$ \\
        $\chi_1$ & first component aligned spins & $0.12$ & $[-0.99, 0.99]$ \\
        $\chi_1$ & second component aligned spins & $0.53$ & $[-0.99, 0.99]$ \\
        $d_L$ & luminosity distance $[\rm{Mpc}]$ & $2500$ & $[500, 4000]$ \\
        $t_c$ & coalescence time $[\rm{s}]$ & $0$ & $[-0.01, 0.01]$ \\
        $\phi_c$ & coalescence phase & $0.4$ & $[0, 2\pi]$ \\
        $\iota$ & inclination angle & $2.5$ & $[0, 2\pi]$ \\
        $\psi$ & polarization angle & $0.4$ & $[0, \pi]$ \\
        $\alpha$ & right ascension & $2.5$ & $[0, 2\pi]$ \\
        $\delta$ & declination & $2.5$ & $[0, 2\pi]$ \\
        \hline \hline
    \end{tabular*}
    \caption{Description of the parameters used in the \ac{GW} simulations, their injected value and uniform prior ranges.}
    \label{tab:parameter_priors}
\end{table*}

\newpage
\section{Corner plot for the 11D example}\label{sec:appendix_11D_fig}
\begin{figure}[h!]
    \centering
    \subfloat[\textsc{Bilby}]{\includegraphics[width=0.65\linewidth]{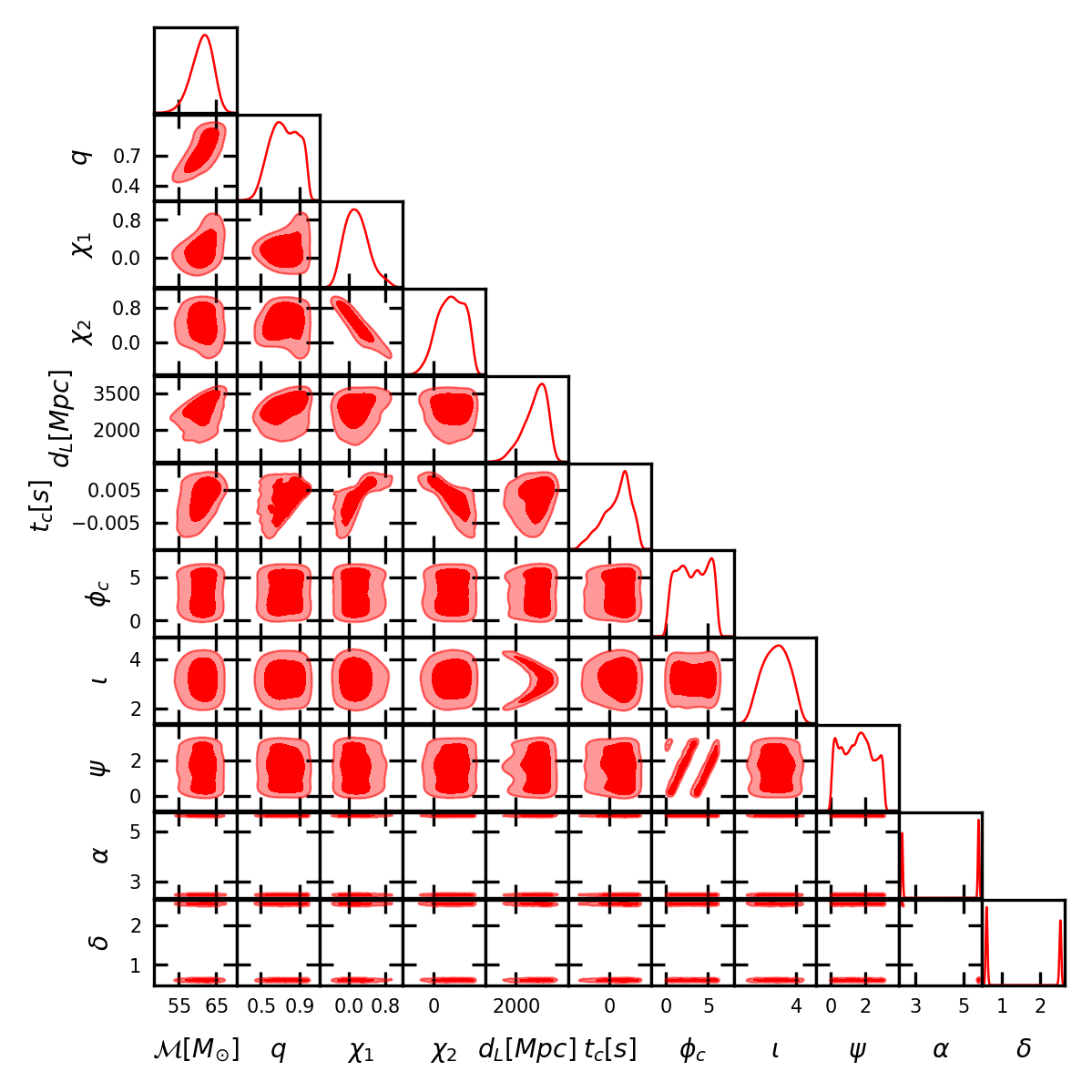}}
    \qquad
    \subfloat[\textsc{Jim} + \texttt{harmonic}]{\includegraphics[width=0.65\linewidth]{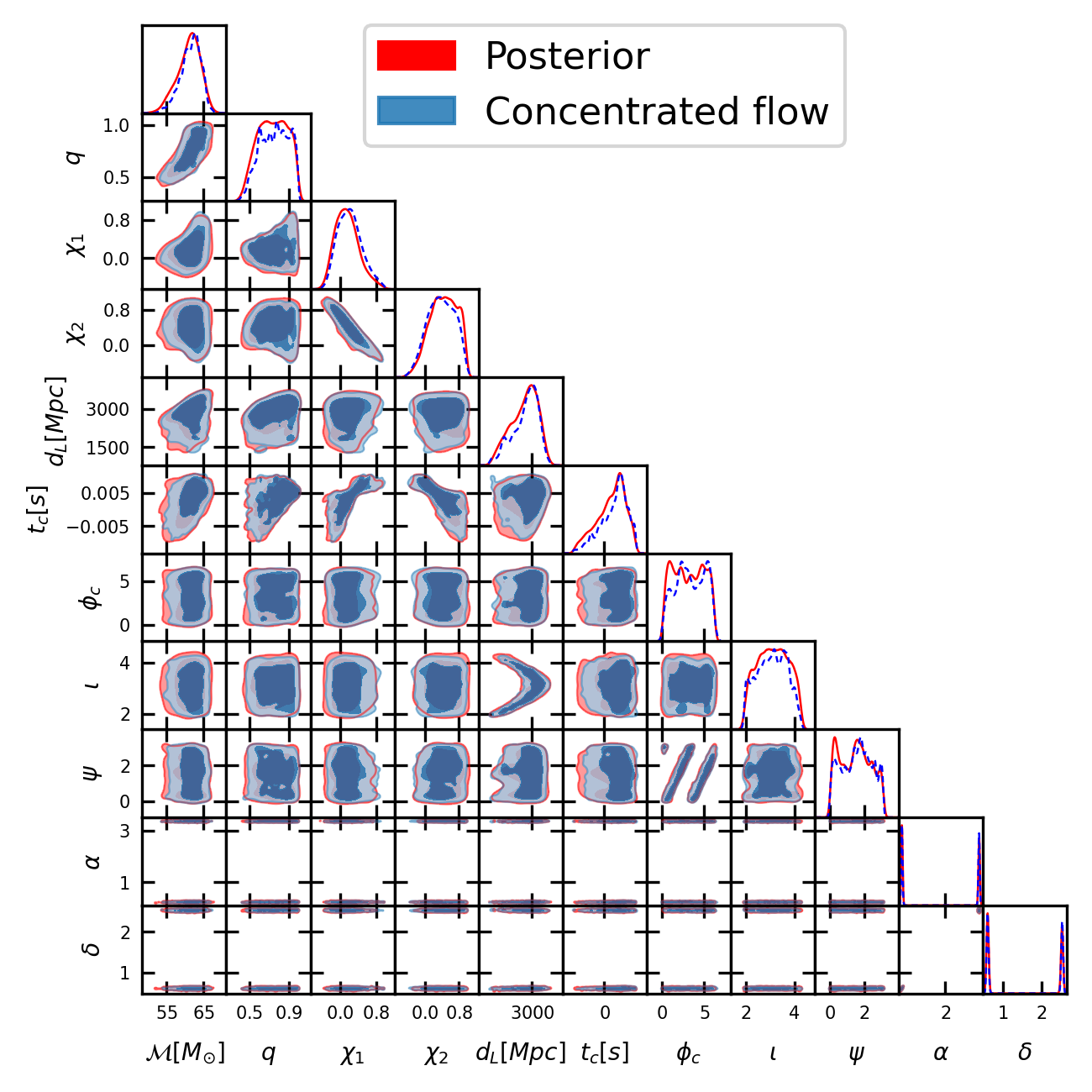}}
    \caption{Corner plots for the 11-dimensional posterior samples from (a) \textsc{Bilby} and (b) \textsc{Jim} used for inference (solid red) alongside the concentrated flow at $T=0.8$ used in the learned harmonic mean (dashed blue).}
    \label{fig:11D}
\end{figure}

\begin{acronym}
    \acro{PE}[PE]{parameter estimation}
    \acro{GW}[GW]{gravitational wave}
    \acrodefplural{GWs}{gravitational waves}
    \acro{NS}[NS]{neutron star}
    \acrodefplural{NSs}{neutron stars}
    \acro{MCMC}[MCMC]{Markov chain Monte Carlo}
    \acro{HMC}[HMC]{Hamiltonian Monte Carlo}
    \acro{MALA}[MALA]{Metropolis-adjusted Langevin algorithm}
    \acro{NF}[NF]{normalizing flow}
    \acro{BBH}[BBH]{binary black hole}
    \acro{BNS}[BNS]{binary neutron star}
    \acro{CPU}[CPU]{central processing unit}
    \acro{GPU}[GPU]{graphical processing unit}
    \acro{TPU}[TPU]{tensor processing unit}
    \acro{ML}[ML]{machine learning}
    \acro{JIT}[JIT]{just-in-time}
    \acro{PSD}[PSD]{power spectral density}
\end{acronym}

\end{document}